\documentclass[%
 reprint,
 amsmath,amssymb,
 aps,
]{revtex4-2}

\usepackage{graphicx}
\usepackage{dcolumn}
\usepackage{bm}
\usepackage{soul}

\usepackage{subcaption}

\begin{document}

\preprint{APS/123-QED}

\title{Alignment Transport Between Ultracold Polar Molecules}
\author{J. Smucker}
 \affiliation{Department of Physics and Astronomy, Stony Brook University, Stony Brook, 11790 USA}
  \email{jonathan.smucker@stonybrook.edu}
\author{J. P\'erez-R\'ios}
 \affiliation{Department of Physics and Astronomy, Stony Brook University, Stony Brook, 11790 USA}

\date{\today}

\begin{abstract}
 We propose an array of ultracold polar molecules as a platform to study alignment transport between molecules. We envision a polar molecule being aligned with an intense off-resonant laser field whose alignment migrates to a nearby molecule due to dipole-dipole interactions. Our results show that the transport of the alignment is due to a complex interplay between electric field-driven excitations and dipole-dipole interactions. All mechanisms for transfer are elucidated and analyzed, and using NaCs, a prototype molecule, we find that the time for alignment transfer is $\mathcal{O}(10\mu\rm s)$, which makes the phenomena readily observable in the lab. 




\end{abstract}

\maketitle

Light-matter interactions are key in many fields of physics, especially in laser induced alignment of molecules~\cite{Stapelfeldt2003} within atomic, molecular and optical physics. When an intense off resonant laser hits a molecule, the molecule will align with the laser polarization axis. In this scenario, there are two competing time scales at play: the rotational period and the pulse duration. When the pulse duration is shorter than the rotational period, entering in the so-called non-adiabatic limit, the alignment experiences revivals which periodicity is controlled by the rotational constant of the molecule~\cite{fleischer2012molecular,Revivals,Revivals2,leibscher2003molecular,horn2006adaptive,alignment,friedrich1995alignment,alignment2}. This effect is seen in isolated molecules as well as in molecules in a quantum solvent, such as helium droplets. In principle, in the case of multiple molecules the alignment could transfer to nearby molecules because of dipole-dipole interactions, which we hypothesize could be observed in arrays of ultracold polar molecules.



Ultracold polar molecules are getting attention due to their prospective applications in quantum information sciences, many-body physics, cold and ultracold chemistry and quantum metrology~\cite{carr2009cold,Krems2004,Krems2008,Bala2016,Quemener2012,Schnell2009,Perez-Rios2020,demille2002quantum}. For instance, arrays of ultracold polar molecules offer a unique scenario to study many-body physics such as exciton dynamics~\cite{perez2010external}, quantum synchronization~\cite{sync}, other many-body phenomena~\cite{many-body1,many-body2,serwatka2024quantum}, and even as a quantum simulator of complex spin Hamiltonians~\cite{quantumsimulator1,quantumsimulator2,quantumsimulator3,quantumsimulator4}. Indeed, alignment transport could be better explored in arrays of ultracold polar molecules. As we will discuss, the transport of the alignment is a complicated process involving multiple intermediate states. This complexity combined with the possibility of expanding to systems of more dipoles and different geometries offer a rich new topic to be explored.




\begin{figure*}[t]
\centering
\includegraphics[scale=0.5]{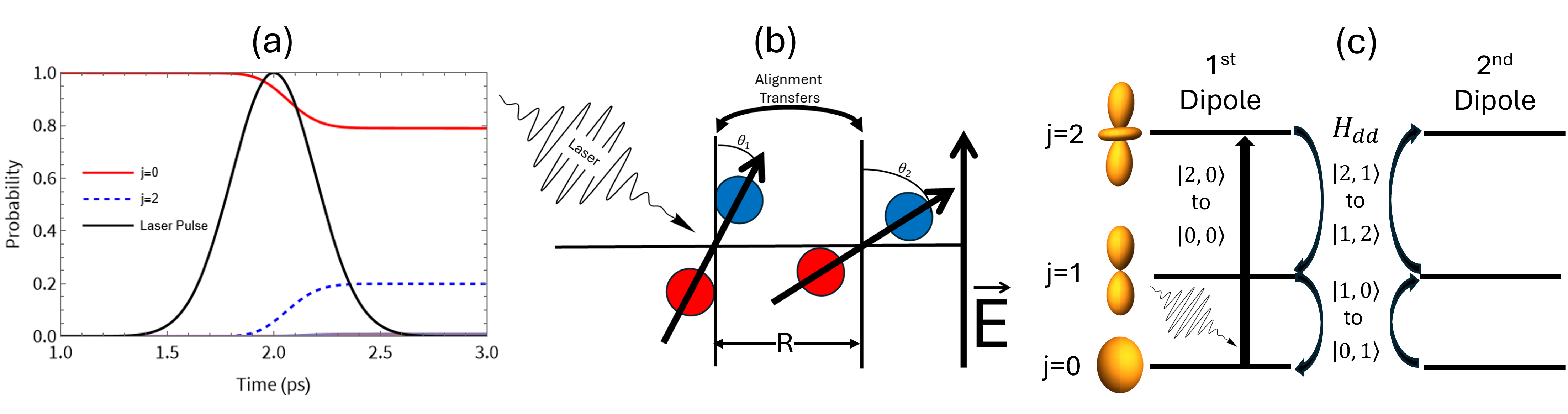}
\caption{Schematic of alignment transport. Fig:~\ref{Schem}a shows the states of the first molecule before and after the laser pulse. Fig:~\ref{Schem}b is a representation of the two molecule system. Fig:~\ref{Schem}c is a sketch of the level diagram of the two dipole system. The laser pulse excites the first molecule to the $j=2$ state and this excitation is passed to the second molecule via dipole-dipole interactions.}
\label{Schem}
\end{figure*}

In this letter, we study alignment transfer between two dipoles in an array assisted by the dipole-dipole interaction, as shown schematically in panel (b) of Fig:~\ref{Schem}. First, a laser pulse excites a molecule, populating the $j=2$ rotational state and inducing the molecules alignment, as shown in panel (a). After this, the presence of a constant electric field plus the dipole-dipole interaction transfers the alignment to the second dipole, as shown in panel (c). Our results show that the transport mechanism is very involved and can be efficiently controlled based on molecular and laser field properties. We will use a combination of techniques to understand the underlying quantum mechanics behind transferring the excitation created by the laser pulse. This understanding will allow us to narrow down the number of states that must be included in our calculations, speeding up our calculations and helping speed up future calculations of larger systems. Many states are required to accurately predict the transport of the alignment because multiple pathways are involved. The effects of the distance between the dipoles, the strength of the dipoles, the rotational constant, and the properties of the laser pulse are all discussed.

In the case of ultracold molecules, the kinetic energy is not enough to excite the vibrational degrees of freedom, and hence it is a good approximation to treat the molecule as a rigid rotor. In this case, the Hamiltonian of the system consists of four different parts:
\begin{equation}
H(t) = H_{1,\text{rot}} + H_{2,\text{rot}} + H_E + H_{dd} + H_{Lsr}(t).
\end{equation}
$H_{i,\text{rot}}$ is the rotational Hamiltonian of the i$^{th}$ molecule. $H_{i,\text{rot}}=B_{0,i}\hbar^2 j_i(j_i+1)$, where $B_{0,i}$ is the i$^{th}$ molecule's equilibrium rotational constant. $H_E=-\bm{\mu}\cdot \bm{E}$ is the Hamiltonian due to the electric field with $\bm{\mu}$ being the dipole of the molecule and $\bm{E}$ being the electric field. $H_{Lsr}$(t) is the Hamiltonian due to the laser pulse. $H_{dd}$ is the Hamiltonian due to the dipole-dipole interaction. We assume that the z-component of the angular momentum $m$ is always zero for both molecules, allowing our states to be represented in Dirac notation as $|j_1,j_2\rangle$ with $j_1$ and $j_2$ representing the total angular momentum quantum numbers for the two dipoles. We also assume our initial state is the $|0,0\rangle$ state.

The dipole-dipole Hamiltonian can be represented as \cite{wei2011entanglement}:
\begin{equation}
    H_{dd}= \frac{\mu^2}{R^3} (1-3\cos^2{\alpha})\cos{\theta_1}\cos{\theta_2},
\end{equation}
assuming the two dipoles have the same dipole moment. $R$ is the distance between the dipoles and $\theta_i$ is the angle of the corresponding $i^{th}$ molecule, shown in Fig:~\ref{Schem}b. We set $\alpha=\pi/2$. The laser pulse's Hamiltonian for a symmetric top or linear molecule is \cite{AlignmentThesisNielsen,friedrich1995alignment}:
\begin{equation}
    H_{Lsr}= -\frac{(E_{Lsr} \epsilon(t))^2}{4}(\alpha_{yy} + \alpha_{\Delta} \cos^2{\theta}),
\end{equation}
where $E_{Lsr}$ is the strength of the laser pulse, $\epsilon(t)$ is the laser pulse profile and $\alpha_{\Delta} = \alpha_{zz} - \alpha_{yy}$. $\alpha_{yy}$ and $\alpha_{zz}$ are the polarizabilities of the molecule being hit by the laser in the $y$ and $z$ directions respectively. Since we are assuming our molecule is symmetric around the z-axis, $\alpha_{yy}=\alpha_{xx}$. Here, we chose NaCs as our molecule with a rotational constant of $0.058$ cm$^{-1}$ \cite{DiatomicDatabasebook,DSCDM}, a polarizability in the parallel direction of $\alpha_{zz}=669.26$ a.u and a polarizability in the perpendicular direction of $\alpha_{yy}=328.25$ a.u. The polarizabilities were calculated using the MP2 level of theory as implemented in the Molpro quantum chemistry package \cite{werner2012molpro}, at the experimental equilibrium distance~\cite{DSCDM} using the AVQZ basis set Na~\cite{AVQZ} and def2-QZVPP for Cs
~\cite{def2weigend2005balanced}. These polarizabilities only impact the effect the laser pulse has on the population of the rotational state and do not impact the alignment transport process. We can ignore the interaction between the laser's electric field and the molecules dipole moment since the laser pulses we consider contain multiple cycles of the electric field causing this interaction to average to approximately zero\cite{AlignmentThesisNielsen,friedrich1995alignment}. We assume a Gaussian pulse: $\epsilon(t)=e^{-2 \log{2} ~t^2/\tau^2}$ with $\tau$ being a time constant that changes the width. For our simulations we set $\tau=1/3$ ps, and hence, we are in the non-adiabatic limit, where revivals in the alignment are expected. We also used a peak laser pulse intensity of $3 \times 10^{13}$ W/cm$^2$ and a constant electric field strength of $0.5$ kV/cm unless otherwise stated.

We delay the peak of the laser pulse by $2$ ps, as shown in panel (a) of Fig.~\ref{Schem}. We solve the time dependent Schr\"{o}dinger equation numerically up to $4$ ps (we set this time to be $t=0$), at this point the laser field is effectively zero. Since $H_{Lsr}(t)$ is the only part of the Hamiltonian that explicitly depends on time, the Hamiltonian can be treated as time independent after $4$ ps allowing for the time dependent Schr\"{o}dinger equation to be solved using separation of variables from this point on. We take the state at $4$ ps from the numerical solution to the Schr\"{o}dinger equation and use it as an initial state for the separation of variables solution. We found that by including $3$ rotational states per molecule, the results are roughly the same as simulations that take into account many more rotational states.

\begin{figure*}
\centering
\includegraphics[scale=0.50]{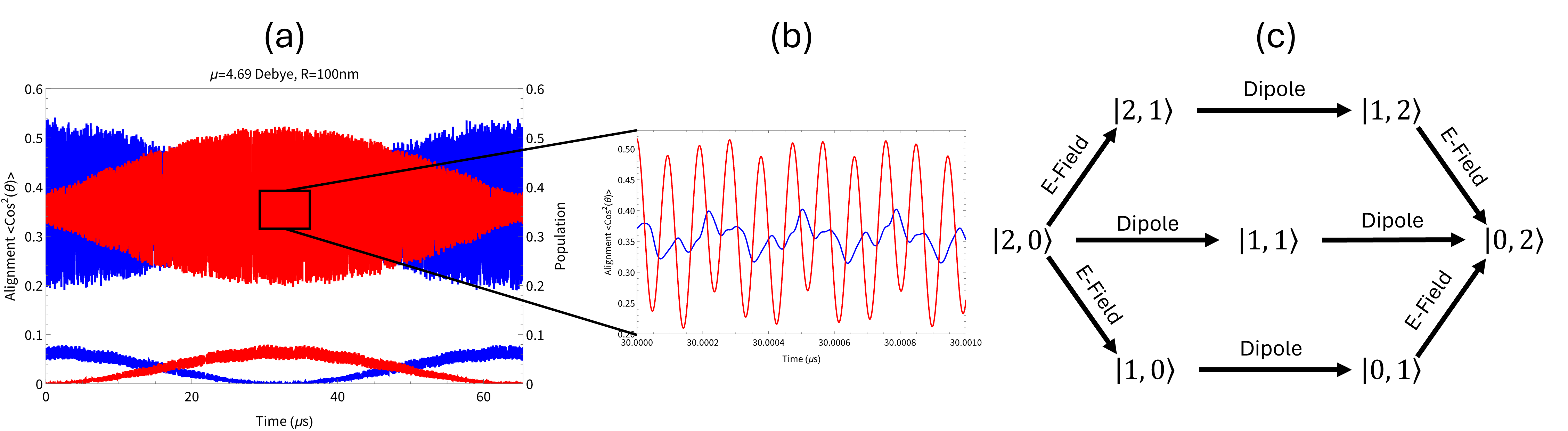}
\caption{The top curves in panel (a) show the alignment as a function of time for two NaCs molecules separated by a distance of $100$ nm. The lower curves in Fig:~\ref{Pathways}a show the population of the $|2,0\rangle$ and the $|0,2\rangle$ states as a function of time. Panel (b) highlights the rapid oscillations of the alignment. Panel (c) shows a schematic view of the most important pathways for the transfer of the population between the $|2,0\rangle$ and the $|0,2\rangle$ states. The text above the arrows show the mechanism allowing for these individual transitions.}
\label{Pathways}
\end{figure*}

Fig:~\ref{Pathways}a shows the result of a simulation with two NaCs molecules separated by a distance of $100$ nm. This figure also shows the population of the $|2,0\rangle$ and the $|0,2\rangle$ states as a function of time. As expected, the alignment on dipole 1 shows revivals. However, due to the presence of the second dipole the lowest alignment of dipole 1 corresponds to the largest alignment of dipole 2, thus, showing how the alignment hops between the two dipoles. The rapid oscillations of the alignment, highlighted in Fig:~\ref{Pathways}b, are mostly due to the energy gap between the $|0,0\rangle$ and the $|2,0\rangle$ states. This short period is therefore around $\pi/3 B_0$ in atomic units.

It is clear from Fig:~\ref{Pathways}a that the transport of the alignment excitation corresponds to a transfer of the population from the $|2,0\rangle$ state to the $|0,2\rangle$ state. These states are not coupled directly via the dipole-dipole interaction, meaning the process must involve at least one intermediate states.

After an extensive study on the role of the rotational states, we found the minimum set of rotational states leading to the same results as using the complete manifold. The relevant states are: $|0,0\rangle$, $|1,0\rangle$, $|2,0\rangle$, $|0,1\rangle$, $|1,1\rangle$, $|2,1\rangle$, $|0,2\rangle$, $|1,2\rangle$, $|2,2\rangle$. Given these states, there are three dominant pathways for the alignment excitation to transfer. These pathways are shown in Fig:~\ref{Pathways}c. The middle pathway in Fig:~\ref{Pathways}c contains only one intermediate state but the dipole-dipole interactions are much slower than the E-field interactions. Therefore, this pathway is less dominant then the other too. Since two of the three pathways depend on the E-field, increasing the E-field will increase the rate of alignment transport slightly. Although, too large of an E-field will mix all the states too efficiently and drastically change the underlying physics.

It is not immediately clear why the $|2,2\rangle$ state is relevant as it does not appear in any of the most relevant pathways. By removing the $|2,2\rangle$ state, the rate of alignment transport increases noticeably. When setting the dipole-dipole interaction between the $|2,2\rangle$ and the $|1,1\rangle$ states to zero, the alignment transport rate does not change. This indicates that it is the E-field interactions with the $|2,1\rangle$ and $|1,2\rangle$ states that make the $|2,2\rangle$ state relevant. Therefore, the $|2,2\rangle$ state slows down the rate of alignment transport by diverting part of the $|2,1\rangle$ and $|1,2\rangle$ states' populations. Similarly, by setting the dipole-dipole interaction between the $|1,1\rangle$ and the $|0,2\rangle$ state to zero (effectively eliminating the pathway shown in the middle of Fig:~\ref{Pathways}c) the rate of alignment transport increases. Again the $|1,1\rangle$ state and the slower middle pathway in Fig:~\ref{Pathways}c divert part of the population from taking the other two faster pathways.
\begin{figure}
\centering
\includegraphics[scale=0.4]{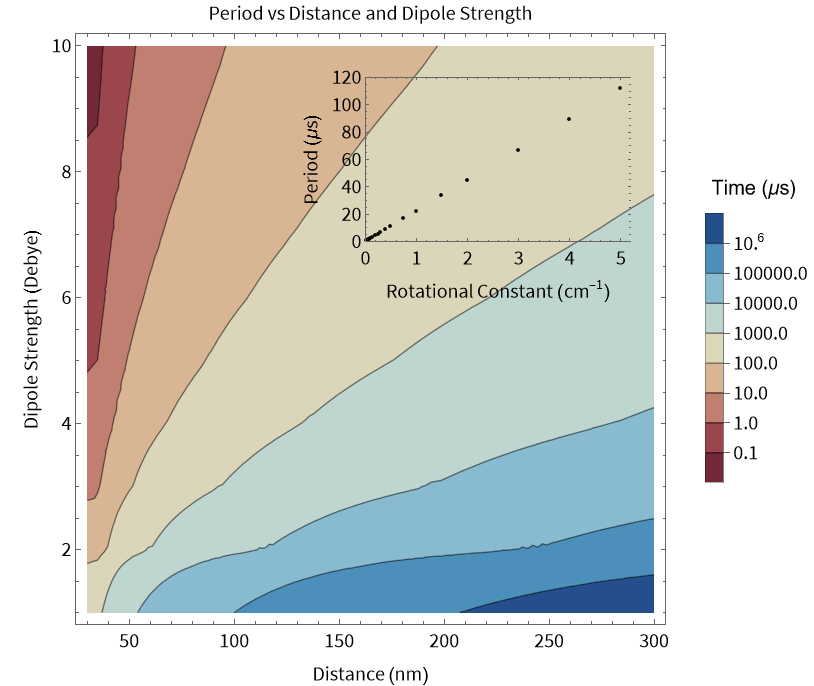}
\caption{The period of the alignment transport as a function of dipole strength and distance. The rotational constant dependence is show in the inset. These calculations were preformed with an E-field strength of $0.25$ kV/cm.}
\label{Period}
\end{figure}

Using Fourier analysis the period of the transport of the alignment can be calculated from our simulations. From this we can find the dipole strength, distance and rotational constant dependence of the period of the alignment excitation. This dependence is shown in Fig:~\ref{Period}. The results for the dipole dependence and distance dependence. shown in Fig:~\ref{Period}, were fit to the function:
\begin{equation}
    T= a \frac{R^3}{\mu^4},
\end{equation}
where $a$ is the fit parameter, $R$ is the distance between the dipoles and $\mu$ is the dipole strength. This function fit to the results from the simulations with an R-Squared value of $0.999957$. This function offered the best fit when compared to functions involving other powers of $R$ and $\mu$. Unsurprisingly, the period depends on $R^3$ as the strength of the dipole-dipole interaction also depends on $R^3$. As we can tell from the relevant pathways shown in Fig:~\ref{Pathways}b, the dipole strength $\mu$ dependence is more complex since both the E-field interactions and the dipole-dipole interactions depend on $\mu$. E-field interactions include a single power of $\mu$ while the dipole-dipole interactions involve $\mu^2$. Since the more dominate pathways (the top pathway and bottom pathways in Fig:~\ref{Pathways}b) both involve two E-field interactions and one dipole-dipole interaction their contributions to the period go as $\mu^4$, which is the period of the alignment transport is proportional to $\mu^{-4}$. The inset in Fig:~\ref{Period} shows the rotational constant dependence which fits to a line with an R-Squared value of $0.999977$.

Up until this point our study has focused on relatively low energy laser pulses. Assuming that the system starts completely in the $|0,0\rangle$ state, when the peak of the pulse is small enough the population remains mostly in the $|0,0\rangle$ state with part of the population being pumped into the $|2,0\rangle$ state. When the energy of the laser pulse is too great the system starts to be pumped into the $|4,0\rangle$ state. The $|4,0\rangle$ state transfers its population to the $|0,4\rangle$ state (through similar multi-state pathways as the $|2,0\rangle$ and $|0,2\rangle$ pathways) but this process takes orders of magnitudes longer. For example, a system of two NaCs molecules separated by a distance of $100$ nm has a predicted period for complete alignment transport of $249$ s if the $|4,0\rangle$ state is populated, compared to the $65$ $\mu$s it takes when the $|4,0\rangle$ state isn't populated. Part of the alignment will still transfer to the observer molecule through the $|2,0\rangle$ and $|0,2\rangle$ pathways with each iteration of this process resulting in more and more of the total alignment excitation being transported. As long as $|4,0\rangle$ state isn't significantly populated the strength of the laser pulse will only affect the magnitude of the alignment and will not affect the period of the alignment transport. The overall shape of the laser pulse should not have an impact on the alignment as long as the same amount of total energy is being pumped into the system. This is true until the laser pulse's duration approaches the period of the short oscillations due to the electric field, i.e., we are in the non-adiabatic regime.

Due to extensive studies of laser cooling and trapping polar molecules\cite{carr2009cold,Schnell2009}, experimental studies of alignment transport should be feasible. Table~\ref{tab:Mols} has the relevant data for a few common polar molecules as well as the predicted period of the alignment transport. The estimated periods are on the same time scale as estimates for CNOT gates using polar molecules \cite{carr2009cold}. The ideal molecule has a large dipole moment and a small rotational constant.

\begin{table}[]
\caption{The periods for alignment transport for various different diatomic molecules. For all simulations the static electric field strength was $0.5$ kV/cm and the peak laser intensity was $3\times 10^{13}$ W/cm$^2$. The distances between the molecules was set to $100$ nm. The dipole strengths, the equilibrium rotational constants and the polarizabilities were predicted from MP2 calculations using AVQZ basis set for Li, Na and K, and the def2-QZVPP basis set for Rb and Cs.}\label{tab:Mols}
\begin{tabular}{cccccc}
\hline
Species & \begin{tabular}[c]{@{}c@{}}$B_0$\\ cm$^{-1}$\end{tabular} & \begin{tabular}[c]{@{}c@{}}$\mu$\\ (Debye)\end{tabular} & \begin{tabular}[c]{@{}c@{}}$\alpha_{zz}$\\ (a.u)\end{tabular} & \begin{tabular}[c]{@{}c@{}}$\alpha_{yy}$\\ (a.u)\end{tabular} & \begin{tabular}[c]{@{}c@{}}Period\\ ($\mu$s)\end{tabular} \\ \hline
NaCs & 0.058 & 4.69 & 669.26 & 328.25 & 65.43 \\ 
LiCs & 0.1907 & 5.5 & 586.41 & 292.93 & 308.38 \\ 
RbCs & 0.015 & 1.24 & 1040.98 & 447.22 & 926.75 \\ 
LiRb & 0.2186 & 4.1 & 500.41 & 266.53 & 1280.82\\ 
KRb & 0.037 & 0.6 & 711.35 & 385.43 & 88021.2 \\ 
\hline
\hline
\end{tabular}
\end{table}

Despite many studies of alignment and of dipole-dipole interactions, studies combining these two effects are minimal. Our results show that ultracold polar molecules can find new applications in the study of alignment transport. Similarly, our findings indicate that the transport time occurs on time scales relevant to quantum gates and can be readily observed in the lab. The relevant states and most important pathways for the alignment transport have been predicted from our simulations. The period of the alignment transport has been determined to depend on the distance cubed and the dipole strength to the negative fourth power. This marks only an introductory study into alignment transport, as a large portion of parameter space is still yet to be explored. Of particular interest is the study of systems involving many dipoles, which drastically increases the parameter space. Preliminary results show that the transport time is reduced when introducing a third dipole in the system, indicating that in an array of ultracold polar molecules, the alignment could efficiently transfer among them. This will be the topic of future studies.

The authors acknowledge the support of the Simons Foundation.

\bibliography{references}

\begin{thebibliography}{33}%
\makeatletter
\providecommand \@ifxundefined [1]{%
 \@ifx{#1\undefined}
}%
\providecommand \@ifnum [1]{%
 \ifnum #1\expandafter \@firstoftwo
 \else \expandafter \@secondoftwo
 \fi
}%
\providecommand \@ifx [1]{%
 \ifx #1\expandafter \@firstoftwo
 \else \expandafter \@secondoftwo
 \fi
}%
\providecommand \natexlab [1]{#1}%
\providecommand \enquote  [1]{``#1''}%
\providecommand \bibnamefont  [1]{#1}%
\providecommand \bibfnamefont [1]{#1}%
\providecommand \citenamefont [1]{#1}%
\providecommand \href@noop [0]{\@secondoftwo}%
\providecommand \href [0]{\begingroup \@sanitize@url \@href}%
\providecommand \@href[1]{\@@startlink{#1}\@@href}%
\providecommand \@@href[1]{\endgroup#1\@@endlink}%
\providecommand \@sanitize@url [0]{\catcode `\\12\catcode `\$12\catcode `\&12\catcode `\#12\catcode `\^12\catcode `\_12\catcode `\%12\relax}%
\providecommand \@@startlink[1]{}%
\providecommand \@@endlink[0]{}%
\providecommand \url  [0]{\begingroup\@sanitize@url \@url }%
\providecommand \@url [1]{\endgroup\@href {#1}{\urlprefix }}%
\providecommand \urlprefix  [0]{URL }%
\providecommand \Eprint [0]{\href }%
\providecommand \doibase [0]{https://doi.org/}%
\providecommand \selectlanguage [0]{\@gobble}%
\providecommand \bibinfo  [0]{\@secondoftwo}%
\providecommand \bibfield  [0]{\@secondoftwo}%
\providecommand \translation [1]{[#1]}%
\providecommand \BibitemOpen [0]{}%
\providecommand \bibitemStop [0]{}%
\providecommand \bibitemNoStop [0]{.\EOS\space}%
\providecommand \EOS [0]{\spacefactor3000\relax}%
\providecommand \BibitemShut  [1]{\csname bibitem#1\endcsname}%
\let\auto@bib@innerbib\@empty
\bibitem [{\citenamefont {Stapelfeldt}\ and\ \citenamefont {Seideman}(2003)}]{Stapelfeldt2003}%
  \BibitemOpen
  \bibfield  {author} {\bibinfo {author} {\bibfnamefont {H.}~\bibnamefont {Stapelfeldt}}\ and\ \bibinfo {author} {\bibfnamefont {T.}~\bibnamefont {Seideman}},\ }\bibfield  {title} {\bibinfo {title} {Colloquium: Aligning molecules with strong laser pulses},\ }\href {https://doi.org/10.1103/RevModPhys.75.543} {\bibfield  {journal} {\bibinfo  {journal} {Rev. Mod. Phys.}\ }\textbf {\bibinfo {volume} {75}},\ \bibinfo {pages} {543} (\bibinfo {year} {2003})}\BibitemShut {NoStop}%
\bibitem [{\citenamefont {Fleischer}\ \emph {et~al.}(2012)\citenamefont {Fleischer}, \citenamefont {Khodorkovsky}, \citenamefont {Gershnabel}, \citenamefont {Prior},\ and\ \citenamefont {Averbukh}}]{fleischer2012molecular}%
  \BibitemOpen
  \bibfield  {author} {\bibinfo {author} {\bibfnamefont {S.}~\bibnamefont {Fleischer}}, \bibinfo {author} {\bibfnamefont {Y.}~\bibnamefont {Khodorkovsky}}, \bibinfo {author} {\bibfnamefont {E.}~\bibnamefont {Gershnabel}}, \bibinfo {author} {\bibfnamefont {Y.}~\bibnamefont {Prior}},\ and\ \bibinfo {author} {\bibfnamefont {I.~S.}\ \bibnamefont {Averbukh}},\ }\bibfield  {title} {\bibinfo {title} {Molecular alignment induced by ultrashort laser pulses and its impact on molecular motion},\ }\href@noop {} {\bibfield  {journal} {\bibinfo  {journal} {Israel Journal of Chemistry}\ }\textbf {\bibinfo {volume} {52}},\ \bibinfo {pages} {414} (\bibinfo {year} {2012})}\BibitemShut {NoStop}%
\bibitem [{\citenamefont {Seideman}(1999)}]{Revivals}%
  \BibitemOpen
  \bibfield  {author} {\bibinfo {author} {\bibfnamefont {T.}~\bibnamefont {Seideman}},\ }\bibfield  {title} {\bibinfo {title} {Revival structure of aligned rotational wave packets},\ }\href {https://doi.org/10.1103/PhysRevLett.83.4971} {\bibfield  {journal} {\bibinfo  {journal} {Phys. Rev. Lett.}\ }\textbf {\bibinfo {volume} {83}},\ \bibinfo {pages} {4971} (\bibinfo {year} {1999})}\BibitemShut {NoStop}%
\bibitem [{\citenamefont {Vrakking}\ \emph {et~al.}(1996)\citenamefont {Vrakking}, \citenamefont {Villeneuve},\ and\ \citenamefont {Stolow}}]{Revivals2}%
  \BibitemOpen
  \bibfield  {author} {\bibinfo {author} {\bibfnamefont {M.~J.~J.}\ \bibnamefont {Vrakking}}, \bibinfo {author} {\bibfnamefont {D.~M.}\ \bibnamefont {Villeneuve}},\ and\ \bibinfo {author} {\bibfnamefont {A.}~\bibnamefont {Stolow}},\ }\bibfield  {title} {\bibinfo {title} {Observation of fractional revivals of a molecular wave packet},\ }\href {https://doi.org/10.1103/PhysRevA.54.R37} {\bibfield  {journal} {\bibinfo  {journal} {Phys. Rev. A}\ }\textbf {\bibinfo {volume} {54}},\ \bibinfo {pages} {R37} (\bibinfo {year} {1996})}\BibitemShut {NoStop}%
\bibitem [{\citenamefont {Leibscher}\ \emph {et~al.}(2003)\citenamefont {Leibscher}, \citenamefont {Averbukh},\ and\ \citenamefont {Rabitz}}]{leibscher2003molecular}%
  \BibitemOpen
  \bibfield  {author} {\bibinfo {author} {\bibfnamefont {M.}~\bibnamefont {Leibscher}}, \bibinfo {author} {\bibfnamefont {I.~S.}\ \bibnamefont {Averbukh}},\ and\ \bibinfo {author} {\bibfnamefont {H.}~\bibnamefont {Rabitz}},\ }\bibfield  {title} {\bibinfo {title} {Molecular alignment by trains of short laser pulses},\ }\href@noop {} {\bibfield  {journal} {\bibinfo  {journal} {Physical review letters}\ }\textbf {\bibinfo {volume} {90}},\ \bibinfo {pages} {213001} (\bibinfo {year} {2003})}\BibitemShut {NoStop}%
\bibitem [{\citenamefont {Horn}\ \emph {et~al.}(2006)\citenamefont {Horn}, \citenamefont {Wollenhaupt}, \citenamefont {Krug}, \citenamefont {Baumert}, \citenamefont {De~Nalda},\ and\ \citenamefont {Banares}}]{horn2006adaptive}%
  \BibitemOpen
  \bibfield  {author} {\bibinfo {author} {\bibfnamefont {C.}~\bibnamefont {Horn}}, \bibinfo {author} {\bibfnamefont {M.}~\bibnamefont {Wollenhaupt}}, \bibinfo {author} {\bibfnamefont {M.}~\bibnamefont {Krug}}, \bibinfo {author} {\bibfnamefont {T.}~\bibnamefont {Baumert}}, \bibinfo {author} {\bibfnamefont {R.}~\bibnamefont {De~Nalda}},\ and\ \bibinfo {author} {\bibfnamefont {L.}~\bibnamefont {Banares}},\ }\bibfield  {title} {\bibinfo {title} {Adaptive control of molecular alignment},\ }\href@noop {} {\bibfield  {journal} {\bibinfo  {journal} {Physical Review A}\ }\textbf {\bibinfo {volume} {73}},\ \bibinfo {pages} {031401} (\bibinfo {year} {2006})}\BibitemShut {NoStop}%
\bibitem [{\citenamefont {Nielsen}\ \emph {et~al.}(2012)\citenamefont {Nielsen}, \citenamefont {Stapelfeldt}, \citenamefont {K\"upper}, \citenamefont {Friedrich}, \citenamefont {Omiste},\ and\ \citenamefont {Gonz\'alez-F\'erez}}]{alignment}%
  \BibitemOpen
  \bibfield  {author} {\bibinfo {author} {\bibfnamefont {J.~H.}\ \bibnamefont {Nielsen}}, \bibinfo {author} {\bibfnamefont {H.}~\bibnamefont {Stapelfeldt}}, \bibinfo {author} {\bibfnamefont {J.}~\bibnamefont {K\"upper}}, \bibinfo {author} {\bibfnamefont {B.}~\bibnamefont {Friedrich}}, \bibinfo {author} {\bibfnamefont {J.~J.}\ \bibnamefont {Omiste}},\ and\ \bibinfo {author} {\bibfnamefont {R.}~\bibnamefont {Gonz\'alez-F\'erez}},\ }\bibfield  {title} {\bibinfo {title} {Making the best of mixed-field orientation of polar molecules: A recipe for achieving adiabatic dynamics in an electrostatic field combined with laser pulses},\ }\href {https://doi.org/10.1103/PhysRevLett.108.193001} {\bibfield  {journal} {\bibinfo  {journal} {Phys. Rev. Lett.}\ }\textbf {\bibinfo {volume} {108}},\ \bibinfo {pages} {193001} (\bibinfo {year} {2012})}\BibitemShut {NoStop}%
\bibitem [{\citenamefont {Friedrich}\ and\ \citenamefont {Herschbach}(1995)}]{friedrich1995alignment}%
  \BibitemOpen
  \bibfield  {author} {\bibinfo {author} {\bibfnamefont {B.}~\bibnamefont {Friedrich}}\ and\ \bibinfo {author} {\bibfnamefont {D.}~\bibnamefont {Herschbach}},\ }\bibfield  {title} {\bibinfo {title} {Alignment and trapping of molecules in intense laser fields},\ }\href@noop {} {\bibfield  {journal} {\bibinfo  {journal} {Physical review letters}\ }\textbf {\bibinfo {volume} {74}},\ \bibinfo {pages} {4623} (\bibinfo {year} {1995})}\BibitemShut {NoStop}%
\bibitem [{\citenamefont {Cai}\ \emph {et~al.}(2001)\citenamefont {Cai}, \citenamefont {Marango},\ and\ \citenamefont {Friedrich}}]{alignment2}%
  \BibitemOpen
  \bibfield  {author} {\bibinfo {author} {\bibfnamefont {L.}~\bibnamefont {Cai}}, \bibinfo {author} {\bibfnamefont {J.}~\bibnamefont {Marango}},\ and\ \bibinfo {author} {\bibfnamefont {B.}~\bibnamefont {Friedrich}},\ }\bibfield  {title} {\bibinfo {title} {Time-dependent alignment and orientation of molecules in combined electrostatic and pulsed nonresonant laser fields},\ }\href {https://doi.org/10.1103/PhysRevLett.86.775} {\bibfield  {journal} {\bibinfo  {journal} {Phys. Rev. Lett.}\ }\textbf {\bibinfo {volume} {86}},\ \bibinfo {pages} {775} (\bibinfo {year} {2001})}\BibitemShut {NoStop}%
\bibitem [{\citenamefont {Carr}\ \emph {et~al.}(2009)\citenamefont {Carr}, \citenamefont {DeMille}, \citenamefont {Krems},\ and\ \citenamefont {Ye}}]{carr2009cold}%
  \BibitemOpen
  \bibfield  {author} {\bibinfo {author} {\bibfnamefont {L.~D.}\ \bibnamefont {Carr}}, \bibinfo {author} {\bibfnamefont {D.}~\bibnamefont {DeMille}}, \bibinfo {author} {\bibfnamefont {R.~V.}\ \bibnamefont {Krems}},\ and\ \bibinfo {author} {\bibfnamefont {J.}~\bibnamefont {Ye}},\ }\bibfield  {title} {\bibinfo {title} {Cold and ultracold molecules: science, technology and applications},\ }\href@noop {} {\bibfield  {journal} {\bibinfo  {journal} {New Journal of Physics}\ }\textbf {\bibinfo {volume} {11}},\ \bibinfo {pages} {055049} (\bibinfo {year} {2009})}\BibitemShut {NoStop}%
\bibitem [{\citenamefont {Doyle}\ \emph {et~al.}(2004)\citenamefont {Doyle}, \citenamefont {Friedrich}, \citenamefont {Krems},\ and\ \citenamefont {Masnou-Seeuws}}]{Krems2004}%
  \BibitemOpen
  \bibfield  {author} {\bibinfo {author} {\bibfnamefont {J.}~\bibnamefont {Doyle}}, \bibinfo {author} {\bibfnamefont {B.}~\bibnamefont {Friedrich}}, \bibinfo {author} {\bibfnamefont {R.~V.}\ \bibnamefont {Krems}},\ and\ \bibinfo {author} {\bibfnamefont {F.}~\bibnamefont {Masnou-Seeuws}},\ }\bibfield  {title} {\bibinfo {title} {Editorial: Quo vadis, cold molecules?},\ }\href {https://doi.org/10.1140/epjd/e2004-00151-x} {\bibfield  {journal} {\bibinfo  {journal} {The European Physical Journal D - Atomic, Molecular, Optical and Plasma Physics}\ }\textbf {\bibinfo {volume} {31}},\ \bibinfo {pages} {149} (\bibinfo {year} {2004})}\BibitemShut {NoStop}%
\bibitem [{\citenamefont {Krems}(2008)}]{Krems2008}%
  \BibitemOpen
  \bibfield  {author} {\bibinfo {author} {\bibfnamefont {R.~V.}\ \bibnamefont {Krems}},\ }\bibfield  {title} {\bibinfo {title} {Cold controlled chemistry},\ }\href {https://doi.org/10.1039/B802322K} {\bibfield  {journal} {\bibinfo  {journal} {Phys. Chem. Chem. Phys.}\ }\textbf {\bibinfo {volume} {10}},\ \bibinfo {pages} {4079} (\bibinfo {year} {2008})}\BibitemShut {NoStop}%
\bibitem [{\citenamefont {Balakrishnan}(2016)}]{Bala2016}%
  \BibitemOpen
  \bibfield  {author} {\bibinfo {author} {\bibfnamefont {N.}~\bibnamefont {Balakrishnan}},\ }\bibfield  {title} {\bibinfo {title} {Perspective: Ultracold molecules and the dawn of cold controlled chemistry},\ }\href {https://doi.org/10.1063/1.4964096} {\bibfield  {journal} {\bibinfo  {journal} {The Journal of Chemical Physics}\ }\textbf {\bibinfo {volume} {145}},\ \bibinfo {pages} {150901} (\bibinfo {year} {2016})},\ \Eprint {https://arxiv.org/abs/https://doi.org/10.1063/1.4964096} {https://doi.org/10.1063/1.4964096} \BibitemShut {NoStop}%
\bibitem [{\citenamefont {Qu{\'e}m{\'e}ner}\ and\ \citenamefont {Julienne}(2012)}]{Quemener2012}%
  \BibitemOpen
  \bibfield  {author} {\bibinfo {author} {\bibfnamefont {G.}~\bibnamefont {Qu{\'e}m{\'e}ner}}\ and\ \bibinfo {author} {\bibfnamefont {P.~S.}\ \bibnamefont {Julienne}},\ }\bibfield  {title} {\bibinfo {title} {Ultracold molecules under control!},\ }\bibfield  {booktitle} {\emph {\bibinfo {booktitle} {Chemical Reviews}},\ }\href {https://doi.org/10.1021/cr300092g} {\bibfield  {journal} {\bibinfo  {journal} {Chemical Reviews}\ }\textbf {\bibinfo {volume} {112}},\ \bibinfo {pages} {4949} (\bibinfo {year} {2012})}\BibitemShut {NoStop}%
\bibitem [{\citenamefont {Schnell}\ and\ \citenamefont {Meijer}(2009)}]{Schnell2009}%
  \BibitemOpen
  \bibfield  {author} {\bibinfo {author} {\bibfnamefont {M.}~\bibnamefont {Schnell}}\ and\ \bibinfo {author} {\bibfnamefont {G.}~\bibnamefont {Meijer}},\ }\bibfield  {title} {\bibinfo {title} {Cold molecules: Preparation, applications, and challenges},\ }\href {https://doi.org/10.1002/anie.200805503} {\bibfield  {journal} {\bibinfo  {journal} {Angewandte Chemie International Edition}\ }\textbf {\bibinfo {volume} {48}},\ \bibinfo {pages} {6010} (\bibinfo {year} {2009})}\BibitemShut {NoStop}%
\bibitem [{\citenamefont {P\'{e}rez-R\'{i}os}(2020)}]{Perez-Rios2020}%
  \BibitemOpen
  \bibfield  {author} {\bibinfo {author} {\bibfnamefont {J.}~\bibnamefont {P\'{e}rez-R\'{i}os}},\ }\href@noop {} {\emph {\bibinfo {title} {An Introduction to Cold and Ultracold Chemistry}}}\ (\bibinfo  {publisher} {Springer International Publishing},\ \bibinfo {address} {Cham, Switzerland},\ \bibinfo {year} {2020})\BibitemShut {NoStop}%
\bibitem [{\citenamefont {DeMille}(2002)}]{demille2002quantum}%
  \BibitemOpen
  \bibfield  {author} {\bibinfo {author} {\bibfnamefont {D.}~\bibnamefont {DeMille}},\ }\bibfield  {title} {\bibinfo {title} {Quantum computation with trapped polar molecules},\ }\href@noop {} {\bibfield  {journal} {\bibinfo  {journal} {Physical Review Letters}\ }\textbf {\bibinfo {volume} {88}},\ \bibinfo {pages} {067901} (\bibinfo {year} {2002})}\BibitemShut {NoStop}%
\bibitem [{\citenamefont {P{\'e}rez-R{\'\i}os}\ \emph {et~al.}(2010)\citenamefont {P{\'e}rez-R{\'\i}os}, \citenamefont {Herrera},\ and\ \citenamefont {Krems}}]{perez2010external}%
  \BibitemOpen
  \bibfield  {author} {\bibinfo {author} {\bibfnamefont {J.}~\bibnamefont {P{\'e}rez-R{\'\i}os}}, \bibinfo {author} {\bibfnamefont {F.}~\bibnamefont {Herrera}},\ and\ \bibinfo {author} {\bibfnamefont {R.~V.}\ \bibnamefont {Krems}},\ }\bibfield  {title} {\bibinfo {title} {External field control of collective spin excitations in an optical lattice of 2$\sigma$ molecules},\ }\href@noop {} {\bibfield  {journal} {\bibinfo  {journal} {New Journal of Physics}\ }\textbf {\bibinfo {volume} {12}},\ \bibinfo {pages} {103007} (\bibinfo {year} {2010})}\BibitemShut {NoStop}%
\bibitem [{\citenamefont {Zhu}\ \emph {et~al.}(2015)\citenamefont {Zhu}, \citenamefont {Schachenmayer}, \citenamefont {Xu}, \citenamefont {Herrera}, \citenamefont {Restrepo}, \citenamefont {Holland},\ and\ \citenamefont {Rey}}]{sync}%
  \BibitemOpen
  \bibfield  {author} {\bibinfo {author} {\bibfnamefont {B.}~\bibnamefont {Zhu}}, \bibinfo {author} {\bibfnamefont {J.}~\bibnamefont {Schachenmayer}}, \bibinfo {author} {\bibfnamefont {M.}~\bibnamefont {Xu}}, \bibinfo {author} {\bibfnamefont {F.}~\bibnamefont {Herrera}}, \bibinfo {author} {\bibfnamefont {J.~G.}\ \bibnamefont {Restrepo}}, \bibinfo {author} {\bibfnamefont {M.~J.}\ \bibnamefont {Holland}},\ and\ \bibinfo {author} {\bibfnamefont {A.~M.}\ \bibnamefont {Rey}},\ }\bibfield  {title} {\bibinfo {title} {Synchronization of interacting quantum dipoles},\ }\href {https://doi.org/10.1088/1367-2630/17/8/083063} {\bibfield  {journal} {\bibinfo  {journal} {New Journal of Physics}\ }\textbf {\bibinfo {volume} {17}},\ \bibinfo {pages} {083063} (\bibinfo {year} {2015})}\BibitemShut {NoStop}%
\bibitem [{\citenamefont {Hazzard}\ \emph {et~al.}(2014)\citenamefont {Hazzard}, \citenamefont {Gadway}, \citenamefont {Foss-Feig}, \citenamefont {Yan}, \citenamefont {Moses}, \citenamefont {Covey}, \citenamefont {Yao}, \citenamefont {Lukin}, \citenamefont {Ye}, \citenamefont {Jin},\ and\ \citenamefont {Rey}}]{many-body1}%
  \BibitemOpen
  \bibfield  {author} {\bibinfo {author} {\bibfnamefont {K.~R.~A.}\ \bibnamefont {Hazzard}}, \bibinfo {author} {\bibfnamefont {B.}~\bibnamefont {Gadway}}, \bibinfo {author} {\bibfnamefont {M.}~\bibnamefont {Foss-Feig}}, \bibinfo {author} {\bibfnamefont {B.}~\bibnamefont {Yan}}, \bibinfo {author} {\bibfnamefont {S.~A.}\ \bibnamefont {Moses}}, \bibinfo {author} {\bibfnamefont {J.~P.}\ \bibnamefont {Covey}}, \bibinfo {author} {\bibfnamefont {N.~Y.}\ \bibnamefont {Yao}}, \bibinfo {author} {\bibfnamefont {M.~D.}\ \bibnamefont {Lukin}}, \bibinfo {author} {\bibfnamefont {J.}~\bibnamefont {Ye}}, \bibinfo {author} {\bibfnamefont {D.~S.}\ \bibnamefont {Jin}},\ and\ \bibinfo {author} {\bibfnamefont {A.~M.}\ \bibnamefont {Rey}},\ }\bibfield  {title} {\bibinfo {title} {Many-body dynamics of dipolar molecules in an optical lattice},\ }\href {https://doi.org/10.1103/PhysRevLett.113.195302} {\bibfield  {journal} {\bibinfo  {journal} {Phys. Rev. Lett.}\ }\textbf {\bibinfo {volume} {113}},\ \bibinfo {pages} {195302} (\bibinfo
  {year} {2014})}\BibitemShut {NoStop}%
\bibitem [{\citenamefont {Capogrosso-Sansone}\ \emph {et~al.}(2010)\citenamefont {Capogrosso-Sansone}, \citenamefont {Trefzger}, \citenamefont {Lewenstein}, \citenamefont {Zoller},\ and\ \citenamefont {Pupillo}}]{many-body2}%
  \BibitemOpen
  \bibfield  {author} {\bibinfo {author} {\bibfnamefont {B.}~\bibnamefont {Capogrosso-Sansone}}, \bibinfo {author} {\bibfnamefont {C.}~\bibnamefont {Trefzger}}, \bibinfo {author} {\bibfnamefont {M.}~\bibnamefont {Lewenstein}}, \bibinfo {author} {\bibfnamefont {P.}~\bibnamefont {Zoller}},\ and\ \bibinfo {author} {\bibfnamefont {G.}~\bibnamefont {Pupillo}},\ }\bibfield  {title} {\bibinfo {title} {Quantum phases of cold polar molecules in 2d optical lattices},\ }\href {https://doi.org/10.1103/PhysRevLett.104.125301} {\bibfield  {journal} {\bibinfo  {journal} {Phys. Rev. Lett.}\ }\textbf {\bibinfo {volume} {104}},\ \bibinfo {pages} {125301} (\bibinfo {year} {2010})}\BibitemShut {NoStop}%
\bibitem [{\citenamefont {Serwatka}\ and\ \citenamefont {Roy}(2024)}]{serwatka2024quantum}%
  \BibitemOpen
  \bibfield  {author} {\bibinfo {author} {\bibfnamefont {T.}~\bibnamefont {Serwatka}}\ and\ \bibinfo {author} {\bibfnamefont {P.-N.}\ \bibnamefont {Roy}},\ }\bibfield  {title} {\bibinfo {title} {Quantum criticality in chains of planar rotors with dipolar interactions},\ }\href@noop {} {\bibfield  {journal} {\bibinfo  {journal} {arXiv preprint arXiv:2401.02887}\ } (\bibinfo {year} {2024})}\BibitemShut {NoStop}%
\bibitem [{\citenamefont {Gorshkov}\ \emph {et~al.}(2011{\natexlab{a}})\citenamefont {Gorshkov}, \citenamefont {Manmana}, \citenamefont {Chen}, \citenamefont {Demler}, \citenamefont {Lukin},\ and\ \citenamefont {Rey}}]{quantumsimulator1}%
  \BibitemOpen
  \bibfield  {author} {\bibinfo {author} {\bibfnamefont {A.~V.}\ \bibnamefont {Gorshkov}}, \bibinfo {author} {\bibfnamefont {S.~R.}\ \bibnamefont {Manmana}}, \bibinfo {author} {\bibfnamefont {G.}~\bibnamefont {Chen}}, \bibinfo {author} {\bibfnamefont {E.}~\bibnamefont {Demler}}, \bibinfo {author} {\bibfnamefont {M.~D.}\ \bibnamefont {Lukin}},\ and\ \bibinfo {author} {\bibfnamefont {A.~M.}\ \bibnamefont {Rey}},\ }\bibfield  {title} {\bibinfo {title} {Quantum magnetism with polar alkali-metal dimers},\ }\href {https://doi.org/10.1103/PhysRevA.84.033619} {\bibfield  {journal} {\bibinfo  {journal} {Phys. Rev. A}\ }\textbf {\bibinfo {volume} {84}},\ \bibinfo {pages} {033619} (\bibinfo {year} {2011}{\natexlab{a}})}\BibitemShut {NoStop}%
\bibitem [{\citenamefont {Gorshkov}\ \emph {et~al.}(2011{\natexlab{b}})\citenamefont {Gorshkov}, \citenamefont {Manmana}, \citenamefont {Chen}, \citenamefont {Ye}, \citenamefont {Demler}, \citenamefont {Lukin},\ and\ \citenamefont {Rey}}]{quantumsimulator2}%
  \BibitemOpen
  \bibfield  {author} {\bibinfo {author} {\bibfnamefont {A.~V.}\ \bibnamefont {Gorshkov}}, \bibinfo {author} {\bibfnamefont {S.~R.}\ \bibnamefont {Manmana}}, \bibinfo {author} {\bibfnamefont {G.}~\bibnamefont {Chen}}, \bibinfo {author} {\bibfnamefont {J.}~\bibnamefont {Ye}}, \bibinfo {author} {\bibfnamefont {E.}~\bibnamefont {Demler}}, \bibinfo {author} {\bibfnamefont {M.~D.}\ \bibnamefont {Lukin}},\ and\ \bibinfo {author} {\bibfnamefont {A.~M.}\ \bibnamefont {Rey}},\ }\bibfield  {title} {\bibinfo {title} {Tunable superfluidity and quantum magnetism with ultracold polar molecules},\ }\href {https://doi.org/10.1103/PhysRevLett.107.115301} {\bibfield  {journal} {\bibinfo  {journal} {Phys. Rev. Lett.}\ }\textbf {\bibinfo {volume} {107}},\ \bibinfo {pages} {115301} (\bibinfo {year} {2011}{\natexlab{b}})}\BibitemShut {NoStop}%
\bibitem [{\citenamefont {Micheli}\ \emph {et~al.}(2006)\citenamefont {Micheli}, \citenamefont {Brennen},\ and\ \citenamefont {Zoller}}]{quantumsimulator3}%
  \BibitemOpen
  \bibfield  {author} {\bibinfo {author} {\bibfnamefont {A.}~\bibnamefont {Micheli}}, \bibinfo {author} {\bibfnamefont {G.~K.}\ \bibnamefont {Brennen}},\ and\ \bibinfo {author} {\bibfnamefont {P.}~\bibnamefont {Zoller}},\ }\bibfield  {title} {\bibinfo {title} {A toolbox for lattice-spin models with polar molecules},\ }\href {https://doi.org/10.1038/nphys287} {\bibfield  {journal} {\bibinfo  {journal} {Nature Physics}\ }\textbf {\bibinfo {volume} {2}},\ \bibinfo {pages} {341} (\bibinfo {year} {2006})}\BibitemShut {NoStop}%
\bibitem [{\citenamefont {Kruckenhauser}\ \emph {et~al.}(2020)\citenamefont {Kruckenhauser}, \citenamefont {Sieberer}, \citenamefont {De~Marco}, \citenamefont {Li}, \citenamefont {Matsuda}, \citenamefont {Tobias}, \citenamefont {Valtolina}, \citenamefont {Ye}, \citenamefont {Rey}, \citenamefont {Baranov},\ and\ \citenamefont {Zoller}}]{quantumsimulator4}%
  \BibitemOpen
  \bibfield  {author} {\bibinfo {author} {\bibfnamefont {A.}~\bibnamefont {Kruckenhauser}}, \bibinfo {author} {\bibfnamefont {L.~M.}\ \bibnamefont {Sieberer}}, \bibinfo {author} {\bibfnamefont {L.}~\bibnamefont {De~Marco}}, \bibinfo {author} {\bibfnamefont {J.-R.}\ \bibnamefont {Li}}, \bibinfo {author} {\bibfnamefont {K.}~\bibnamefont {Matsuda}}, \bibinfo {author} {\bibfnamefont {W.~G.}\ \bibnamefont {Tobias}}, \bibinfo {author} {\bibfnamefont {G.}~\bibnamefont {Valtolina}}, \bibinfo {author} {\bibfnamefont {J.}~\bibnamefont {Ye}}, \bibinfo {author} {\bibfnamefont {A.~M.}\ \bibnamefont {Rey}}, \bibinfo {author} {\bibfnamefont {M.~A.}\ \bibnamefont {Baranov}},\ and\ \bibinfo {author} {\bibfnamefont {P.}~\bibnamefont {Zoller}},\ }\bibfield  {title} {\bibinfo {title} {Quantum many-body physics with ultracold polar molecules: Nanostructured potential barriers and interactions},\ }\href {https://doi.org/10.1103/PhysRevA.102.023320} {\bibfield  {journal} {\bibinfo  {journal} {Phys. Rev. A}\ }\textbf {\bibinfo
  {volume} {102}},\ \bibinfo {pages} {023320} (\bibinfo {year} {2020})}\BibitemShut {NoStop}%
\bibitem [{\citenamefont {Wei}\ \emph {et~al.}(2011)\citenamefont {Wei}, \citenamefont {Kais}, \citenamefont {Friedrich},\ and\ \citenamefont {Herschbach}}]{wei2011entanglement}%
  \BibitemOpen
  \bibfield  {author} {\bibinfo {author} {\bibfnamefont {Q.}~\bibnamefont {Wei}}, \bibinfo {author} {\bibfnamefont {S.}~\bibnamefont {Kais}}, \bibinfo {author} {\bibfnamefont {B.}~\bibnamefont {Friedrich}},\ and\ \bibinfo {author} {\bibfnamefont {D.}~\bibnamefont {Herschbach}},\ }\bibfield  {title} {\bibinfo {title} {Entanglement of polar molecules in pendular states},\ }\href@noop {} {\bibfield  {journal} {\bibinfo  {journal} {The Journal of chemical physics}\ }\textbf {\bibinfo {volume} {134}} (\bibinfo {year} {2011})}\BibitemShut {NoStop}%
\bibitem [{\citenamefont {Nielsen}(2012)}]{AlignmentThesisNielsen}%
  \BibitemOpen
  \bibfield  {author} {\bibinfo {author} {\bibfnamefont {J.~H.}\ \bibnamefont {Nielsen}},\ }\emph {\bibinfo {title} {Laser-induced alignment and orientation of quantum-state selected molecules and molecules in liquid helium droplets}},\ \href@noop {} {Ph.D. thesis},\ \bibinfo  {school} {Aarhus University} (\bibinfo {year} {2012})\BibitemShut {NoStop}%
\bibitem [{\citenamefont {Herzberg}(1945)}]{DiatomicDatabasebook}%
  \BibitemOpen
  \bibfield  {author} {\bibinfo {author} {\bibfnamefont {G.}~\bibnamefont {Herzberg}},\ }\href@noop {} {\emph {\bibinfo {title} {Molecular spectra and molecular structure}}}\ (\bibinfo  {publisher} {D. van Nostrand},\ \bibinfo {year} {1945})\BibitemShut {NoStop}%
\bibitem [{\citenamefont {Wang}\ \emph {et~al.}(2023)\citenamefont {Wang}, \citenamefont {Julian}, \citenamefont {Ibrahim}, \citenamefont {Chin}, \citenamefont {Bhattiprolu}, \citenamefont {Franco},\ and\ \citenamefont {P{\'e}rez-R{\'\i}os}}]{DSCDM}%
  \BibitemOpen
  \bibfield  {author} {\bibinfo {author} {\bibfnamefont {Y.}~\bibnamefont {Wang}}, \bibinfo {author} {\bibfnamefont {D.}~\bibnamefont {Julian}}, \bibinfo {author} {\bibfnamefont {M.~A.}\ \bibnamefont {Ibrahim}}, \bibinfo {author} {\bibfnamefont {C.}~\bibnamefont {Chin}}, \bibinfo {author} {\bibfnamefont {S.}~\bibnamefont {Bhattiprolu}}, \bibinfo {author} {\bibfnamefont {E.}~\bibnamefont {Franco}},\ and\ \bibinfo {author} {\bibfnamefont {J.}~\bibnamefont {P{\'e}rez-R{\'\i}os}},\ }\bibfield  {title} {\bibinfo {title} {The database of spectroscopic constants of diatomic molecules (dscdm): A dynamic and user-friendly interface for molecular physics and spectroscopy},\ }\href {https://doi.org/https://doi.org/10.1016/j.jms.2023.111848} {\bibfield  {journal} {\bibinfo  {journal} {Journal of Molecular Spectroscopy}\ }\textbf {\bibinfo {volume} {398}},\ \bibinfo {pages} {111848} (\bibinfo {year} {2023})}\BibitemShut {NoStop}%
\bibitem [{\citenamefont {W.}\ \emph {et~al.}(2012)\citenamefont {W.}, \citenamefont {Knowles}, \citenamefont {Knizia}, \citenamefont {Manby},\ and\ \citenamefont {Sch{\"u}tz}}]{werner2012molpro}%
  \BibitemOpen
  \bibfield  {author} {\bibinfo {author} {\bibfnamefont {H.-J.}\ \bibnamefont {W.}}, \bibinfo {author} {\bibfnamefont {P.~J.}\ \bibnamefont {Knowles}}, \bibinfo {author} {\bibfnamefont {G.}~\bibnamefont {Knizia}}, \bibinfo {author} {\bibfnamefont {F.~R.}\ \bibnamefont {Manby}},\ and\ \bibinfo {author} {\bibfnamefont {M.}~\bibnamefont {Sch{\"u}tz}},\ }\bibfield  {title} {\bibinfo {title} {Molpro: a general-purpose quantum chemistry program package},\ }\href@noop {} {\bibfield  {journal} {\bibinfo  {journal} {Wiley Interdisciplinary Reviews: Computational Molecular Science}\ }\textbf {\bibinfo {volume} {2}},\ \bibinfo {pages} {242} (\bibinfo {year} {2012})}\BibitemShut {NoStop}%
\bibitem [{\citenamefont {Prascher}\ \emph {et~al.}(2011)\citenamefont {Prascher}, \citenamefont {Woon}, \citenamefont {Peterson}, \citenamefont {Dunning},\ and\ \citenamefont {Wilson}}]{AVQZ}%
  \BibitemOpen
  \bibfield  {author} {\bibinfo {author} {\bibfnamefont {B.~P.}\ \bibnamefont {Prascher}}, \bibinfo {author} {\bibfnamefont {D.~E.}\ \bibnamefont {Woon}}, \bibinfo {author} {\bibfnamefont {K.~A.}\ \bibnamefont {Peterson}}, \bibinfo {author} {\bibfnamefont {T.~H.}\ \bibnamefont {Dunning}},\ and\ \bibinfo {author} {\bibfnamefont {A.~K.}\ \bibnamefont {Wilson}},\ }\bibfield  {title} {\bibinfo {title} {Gaussian basis sets for use in correlated molecular calculations. vii. valence, core-valence, and scalar relativistic basis sets for li, be, na, and mg},\ }\href {https://doi.org/10.1007/s00214-010-0764-0} {\bibfield  {journal} {\bibinfo  {journal} {Theoretical Chemistry Accounts}\ }\textbf {\bibinfo {volume} {128}},\ \bibinfo {pages} {69} (\bibinfo {year} {2011})}\BibitemShut {NoStop}%
\bibitem [{\citenamefont {Weigend}\ and\ \citenamefont {Ahlrichs}(2005)}]{def2weigend2005balanced}%
  \BibitemOpen
  \bibfield  {author} {\bibinfo {author} {\bibfnamefont {F.}~\bibnamefont {Weigend}}\ and\ \bibinfo {author} {\bibfnamefont {R.}~\bibnamefont {Ahlrichs}},\ }\bibfield  {title} {\bibinfo {title} {Balanced basis sets of split valence, triple zeta valence and quadruple zeta valence quality for h to rn: Design and assessment of accuracy},\ }\href@noop {} {\bibfield  {journal} {\bibinfo  {journal} {Physical Chemistry Chemical Physics}\ }\textbf {\bibinfo {volume} {7}},\ \bibinfo {pages} {3297} (\bibinfo {year} {2005})}\BibitemShut {NoStop}%
\end{thebibliography}%

\end{document}